\documentclass[dvipdfmx,hiresbb]{pasj01}
\draft

\Received{2018/10/03}
\Accepted{2018/11/21}
 
\newcommand{\argmin}{\mathop{\rm arg~min}\limits}

\begin{document} 

\title{ 
An Image Reconstruction Method for the X-ray Telescope System
with an Angular Resolution Booster}

\author{Mikio \textsc{Morii}\altaffilmark{1}%
}
\altaffiltext{1}{Research Center for Statistical Machine Learning,
The Institute of Statistical Mathematics,
10-3 Midori-cho, Tachikawa, Tokyo 190-8562, Japan}
\email{morii@ism.ac.jp}

\author{Shiro \textsc{Ikeda},\altaffilmark{1}}

\author{Yoshitomo \textsc{Maeda}\altaffilmark{2}}
\altaffiltext{2}{Institute of Space and Astronautical Science, 
Japan Aerospace Exploration Agency,
3-1-1 Yoshinodai, Chuo-ku, Sagamihara, Kanagawa 252-5210, Japan}


\maketitle

\begin{abstract}
We propose an image reconstruction method for an X-ray telescope system
with an angular resolution booster proposed by \citet{Maeda+2018}.
The system consists of double multi-grid masks in front of an X-ray mirror
and an off-focused two-dimensional imager.
Because the obtained image is off-focused, additional image reconstruction process
is assumed to be included.
Our image reconstruction method is 
an extension of the traditional Richardson-Lucy algorithm
with two regularization terms, one for sparseness and the other for smoothness.
Such a combination is desirable for astronomical imaging 
because astronomical objects have variety in shape from point sources, 
diffuse sources to mixtures of them.
The performance of the system is demonstrated with simulated data
for point sources and diffused X-ray sources such as Cas A and Crab Nebula.
The image resolution is improved from a few arcmin of focused image without the booster
to a few arcsec with the booster.
Through the demonstration, the angular resolution booster
with the image reconstruction method is shown to be feasible.
\end{abstract}

\section{Introduction}

Among X-ray telescopes, {\it Chandra} X-ray satellite \citep{Weisskopf+2000} attains
the best angular resolution ($0.5$ arcsec).
However, the effective area of the focusing mirror is not large enough to collect
photons for statistical study of time variation and energy spectrum.
X-ray telescopes like {\it Suzaku} \citep{Mitsuda+2007}
and {\it Hitomi} \citep{Takahashi+2016} were designed
to have moderately high angular resolution of a few arcmin and large effective area.
If the angular resolution of {\it Suzaku}/{\it Hitomi} telescopes 
had been improved to a few arcsec preserving the effective area,
different scientific topics would be revealed. Two examples are shown below.



The first example is the lensed quasar.
\citet{Chartas+2017} and references therein monitored several lensed quasars
in multi-wavelength including the {\it Chandra} X-ray measurements.
They found red-shifted and blue-shifted Fe K$\alpha$ lines in the spectra of a lensed images.
They interpreted the shift of the Fe K$\alpha$ line as resulting
from gravitational and special relativistic Doppler effects.
Since the image separation of lensed quasars is generally larger than a few tenths of arcsec,
often down to a few arcsecs, the larger effective area with an angular resolution of arcsecs
becomes a unique probe of lensed quasars.
The other example is the study of supernova remnants.
\citet{Uchiyama_Aharonian_2008, Sato+2018} revealed
the mechanism of particle acceleration in supernova remnants 
by observing time variation of the shell in arcsec resolution
for a bright supernova remnant Cas A \citep{Hughes+2000} with {\it Chandra}.
More of such studies are desired by telescopes with larger effective area than {\it Chandra}.


Recently, our companion paper \citep{Maeda+2018} has proposed
a simple angular resolution booster for an X-ray telescope like {\it Hitomi},
which can improve the angular resolution by order of 1 - 2 possibly with low-cost.
The system consists of a set of multi-grid masks in front of
the X-ray mirror and two-dimensional imager at a position slightly off-focused
from the focal plane.
The idea is to utilize the shadow pattern shed on the imager through the masks
which contains the information of angular resolution higher than that obtained only with the mirror.
Since the imager's position is off-focused,
the system needs to reconstruct the sky image from the observed pattern.

For the image reconstruction, the Richardson-Lucy algorithm
\citep{Richardson_1972, Lucy_1974} is well-known
as a method to obtain the sky image from the data following a Poisson statistics.
\citet{Ikeda+2014} modified the method with a regularizer of sparseness.
In this paper, we show a new image reconstruction method
by extending these methods with
two regularization terms, where one is for sparseness and the other is for smoothness.

\citet{Starck_Pantin_Murtagh_2002} and \citet{Puetter_Gosnell_Yahil_2005}
has reviewed image deconvolution methods using Fourier or Wavelet basis
for improvement of image resolution, which are different approach from our work.

\section{Mathematical formulation of the problem}

In this article, the image is an $M = m\times n$ pixel rectangular region.
This corresponds to a rectangular area on the tangential plane of the celestial sphere.
Each pixel is indexed with $u = (i, j)$ ($i = 1, \cdots, m; j = 1, \cdots, n$).
The astronomical image is expressed by the distribution of intensity of photons
per unit dimension (for example; per unit sky area per unit time, deg$^{-2}$ sec$^{-1}$,
or per unit sky area per unit time per unit energy range, deg$^{-2}$ sec$^{-1}$ erg$^{-1}$, etc).
Here, we express astronomical image $I(u)$ at each pixel $u$ as a non-negative real value.
On the other hand, pixels of off-focus plane detector is indexed with $v$ ($v = 1, \cdots, V$).
The number of photons in an exposure (an observation) of the telescope detected at a pixel $v$ is $Y(v)$,
which is a non-negative integer value.
It follows a Poisson distribution:
$Y(v) \sim {\rm Poisson} \left( \sum_u t(v, u) I(u) \right)$,
where $t(v, u)$ is the response of the detector
(including the telescope and imager responses) from a pixel $u$ with a unit intensity.
We assume each element of the response matrix is a non-negative value and normalized by $\sum_v t(v, u) = 1$.
The normalized image $\rho(u)$ is expressed by $\rho(u) = \frac{I(u)}{S}$, where $S = \sum_u I(u)$.

The joint distribution of $Y_v$ is described as a product of two probability functions. 
One is the distribution of $\sum_v Y_v$, which follows the Poisson distribution
${\rm Poisson} \left(\sum_v Y_v ; S \right)$ and the other is that of $Y_v$,
which follows the multinomial distribution
${\rm Multinomial} \left(\{Y_v\}; \{\sum_u t(v, u) \rho_u \}\right)$.
\begin{eqnarray}
  & & P(Y | I) = \prod_v {\rm Poisson} \left( Y_v ; \sum_u t(v, u) I(u) \right) \nonumber \\
  & = & \frac{e^{-S} S^{(\sum_v Y_v)}}{(\sum_v Y_v)!} \times
  \frac{(\sum_v Y_v)!}{\prod_v Y_v!} \prod_v \left[ \sum_u t(v, u) \rho_u \right]^{Y_v} \\
  & = & {\rm Poisson} \left(\sum_v Y_v ; S \right) \times
       {\rm Multinomial} \left(\{Y_v\}; \Bigl\{\sum_u t(v, u) \rho_u \Bigr\}\right).
\end{eqnarray}
The above equation shows $S$ and $\rho$ can be estimated separately.
A natural estimate of $S$ is the maximum likelihood estimator, 
which is easily obtained as $S$: $S^* = \sum_v Y(v)$.
So, from now on, we consider only the likelihood function of $\rho$.
By taking the logarithm and neglecting the terms independent of $\rho$, 
the likelihood function for $\rho$ becomes
$L_\rho(\rho) = \sum_{v} Y(v) \log \left[ \sum_u t(v, u) \rho(u) \right].$

It is known that the maximum likelihood of $L_\rho(\rho)$ is calculated 
by the Richardson-Lucy algorithm \citep{Richardson_1972, Lucy_1974}.
However, when the number of the photons is small, the estimated image becomes unstable. 
This is because the information required to reconstruct the $m\times n$ image is
much larger compared to the limited photon counts.

In order to overcome the problem, we propose a method based on sparse modeling. 
Sparse modeling is a new signal processing method. Assuming a sparseness of the image in some domain, 
the amount of information to be estimated is reduced. 
In this case, we assume the image $\rho$ is sparse (many pixels are zeros) and smooth. 
Under this assumption, the image would be reconstructed from limited photons. 
These two types of assumptions is suitable for astronomical imaging,
because astronomical objects have variety in shape from point sources,
diffuse sources (e.g. supernova remnants, clusters of galaxies, and pulsar wind nebula)
to mixtures of them (e.g. point sources in Galactic planes).

Common implementation of the sparse modeling is to use additional regularization terms for the optimization. 
In our case, we employ a term to encourage the sparseness and another term for smoothness. 
Widely used regularization term for sparseness is the $\ell_1$ norm, 
but it does not work in this case since $\|\rho(u)\|_1 = \sum_u\rho(u) = 1$ from its definition. 
Thus, we follow the idea in \citet{Ikeda+2014}, which uses the following term
\begin{equation}
  (1-\beta) \sum_u\log\rho(u), \hspace{1em} 0< \beta.
\end{equation}
Here, the parameter $\beta (> 0)$ controls the sparseness. 
Smaller $\beta$ produces sparser image, and the term disappear when $\beta = 1$. 
We call this the Dirichlet term, since it is related to the Dirichlet distribution.

For the regularization of smoothness, we add the following regularization term in the log-likelihood,
which works to decrease the difference of pixel values between adjoining pixels:
\begin{eqnarray}
  V({\rho}) & = & \sum_{i = 1}^{m - 1} \sum_{j = 1}^{n - 1}
  \left[ (\rho_{i,j} - \rho_{i + 1,j})^2
    + (\rho_{i,j} - \rho_{i,j + 1})^2 
    \right] \\
  & & + \sum_{i = 1}^{m - 1} (\rho_{i,n} - \rho_{i + 1,n})^2
  + \sum_{j = 1}^{n - 1} (\rho_{m,j} - \rho_{m, j+1})^2 .
\end{eqnarray}
Finally, the cost function of $\rho$ to be minimized becomes
\begin{equation}
  L(\rho) = - L_\rho(\rho) + (1 - \beta) \sum_u \log \rho(u) + \mu V(\rho),
\end{equation}
where the hyper parameters $\beta$ and $\mu$ control the degree of
sparseness and smoothness, respectively.
We optimize $\rho$ in the valid region for sky images:
$C = \{\rho \in R^M | \rho_u \geq 0, \sum_u \rho_u = 1\}$.

The optimization problem of maximizing $L_\rho(\rho)$ is known to be solved by the Richardson-Lucy algorithm
\citep{Richardson_1972, Lucy_1974},
which is indeed the EM algorithm for the distribution obeying Poisson statistics \citep{Dempster+1977}.
The problem of maximizing $L_\rho(\rho) - (1 - \beta) \sum_u \log \rho(u)$ is also solved with
the EM algorithm as shown in \citet{Ikeda+2014}.
The current problem cannot be solved only with the EM algorithm,
then we combined the EM algorithm and the proximal gradient method
in order to solve it (e.g. \citet{Beck-Teboulle_2009}).

\section{Algorithm for the reconstruction}

At first, we apply the EM algorithm for the optimization of the cost function,
then the $r$-th step in the EM M-step becomes
\begin{eqnarray}
  \rho^{(r + 1)} & = & \argmin_{\rho \in C}
  \left[ L_{\rm sub} \equiv - \sum_u m_u^{(r)} \log \rho_u + (1 - \beta) \sum_u \log \rho_u + \mu V(\rho) \right],
\end{eqnarray}
where
\begin{equation}
  m_u^{(r)} = \sum_{v} Y(v) \frac{ t(v, u) \rho_u^{(r)}}{ \sum_{u} t(v, u) \rho_u^{(r)} }
\end{equation}
($r = 1, 2, \cdots$).
To solve the sub-problem, we use the proximal gradient method as shown in \citet{Beck-Teboulle_2009}.
Here, we divide $L_{\rm sub}$ into two terms $L_{\rm sub} = g(\rho) + f(\rho)$,
where $f(\rho) = \mu V(\rho)$ and $g(\rho) = - \sum_u [m_u^{(r)} - (1 - \beta)] \log \rho_u$.
We follow the algorithm of ``ISTA with backtracking,''
then the updating rule of the proximal gradient method in the $k$-th step becomes
\begin{eqnarray}
  \rho^{(k + 1)} & = & \argmin_{\rho \in C}
  \sum_u \left[ \frac{L_k}{2} (\rho_u - \sigma_u^{(k)})^2 - [m_u^{(r)} - (1 - \beta)] \log \rho_u \right],
  \label{eq: proxmap}
\end{eqnarray}
where
\begin{equation}
  \sigma_u^{(k)} = \rho_u^{(k)}
  - \frac{\mu}{L_k} \frac{\partial V(\rho^{(k)})}{\partial \rho_u}
\end{equation}
($k = 1, 2, \cdots$).
Here, the Lipschitz constant $L_k$ is searched for in each $k$-step.
This minimization problem can be solved in a closed form by introducing a Lagrange multiplier.
The details are shown in Appendix.
Then, we can solve this minimization problem for a fixed ($\beta$, $\mu$).
Since the reconstructed image depends on the hyper parameters ($\beta$, $\mu$),
we determine the best parameters by the cross-validation.


\section{Results}

\begin{figure}
 \begin{center}
   \includegraphics[width=15cm]{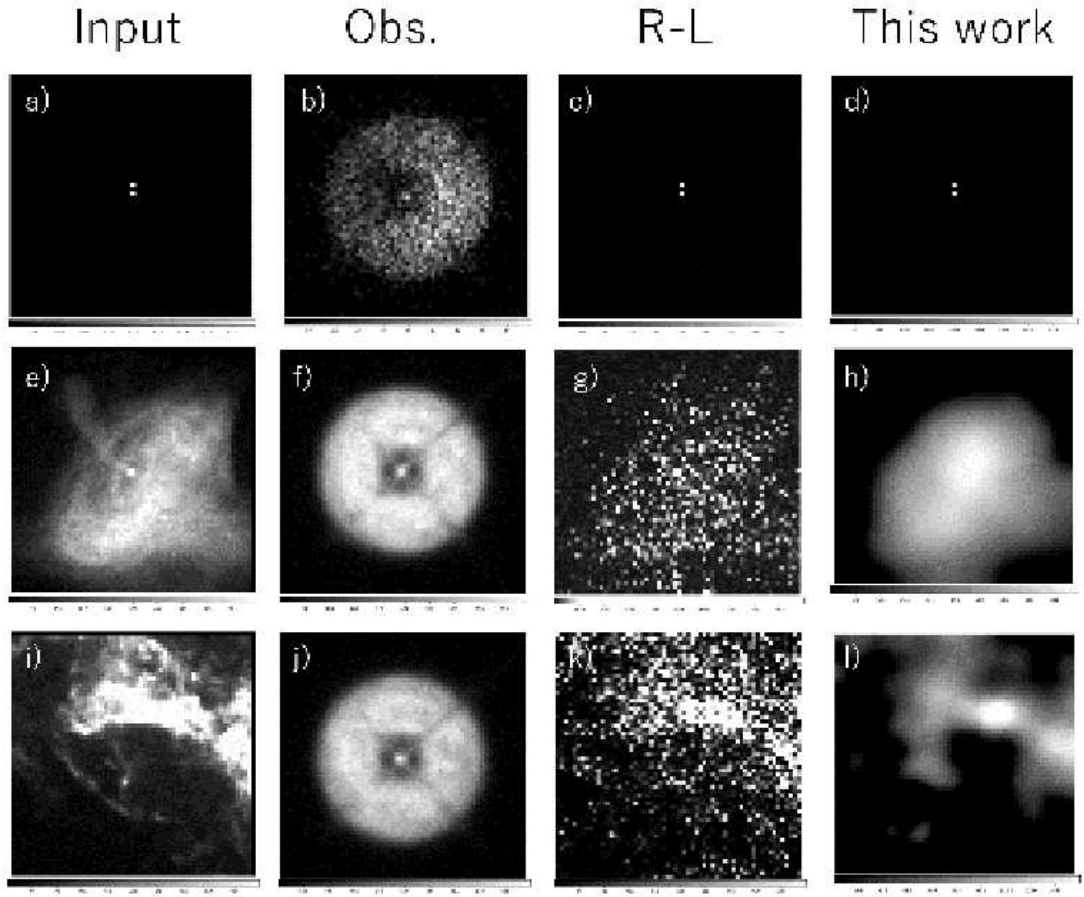}
 \end{center}
\caption{The image reconstruction for double stars (top), Crab nebula (middle) and Cas A (bottom).
The input images are shown in the first column. The panel (a) is double stars with a separation of 4 arcsec.
The panels (e, i) are images of Crab nebula and Cas A obtained with {\it Chandra} X-ray satellite, respectively.
The second columns are simulated observed images on the off-focus imager.
The number of photons from the two stars is $5 \times 10^3$ photons each (b), and the total number of photons 
of Crab nebula and Cas A are $10^6$ (f, j). The reconstructed images by the Richardson-Lucy method and our proposed
method are shown in the third and fourth columns, respectively.
The images (d, h, l) are the results with the best hyper parameters,
$(\beta, \log_{10} \mu) = (0.6, 1.0)$, $(0.6, 8.0)$ and $(0.9, 7.0)$, respectively.
The pixel scale of the input and reconstructed images are 2 arcsec per one pixel.
The images above and elsewhere are made in the look down geometry.
}\label{fig:reconst}
\end{figure}

\begin{figure}
 \begin{center}
   \includegraphics[width=8cm]{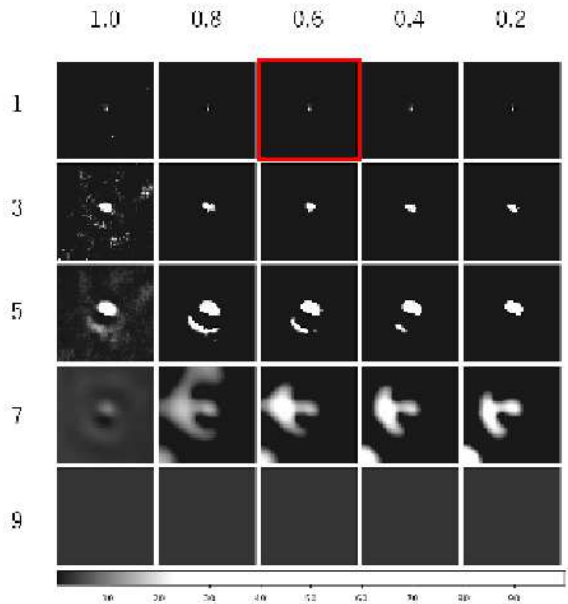}
 \end{center}
\caption{The reconstructed double stars image by changing hyper parameters ($\beta$, $\log_{10} \mu$).
The sparsity and smoothness increase from the left to the right and
from the top to the bottom, by setting $\beta$ and $\log_{10} \mu$ as shown 
above and left of the figure, respectively.
The best image determined by the cross-validation is marked with a red frame.}\label{fig:double_star_mu_beta}
\end{figure}

\begin{figure}
 \begin{center}
   \includegraphics[width=8cm]{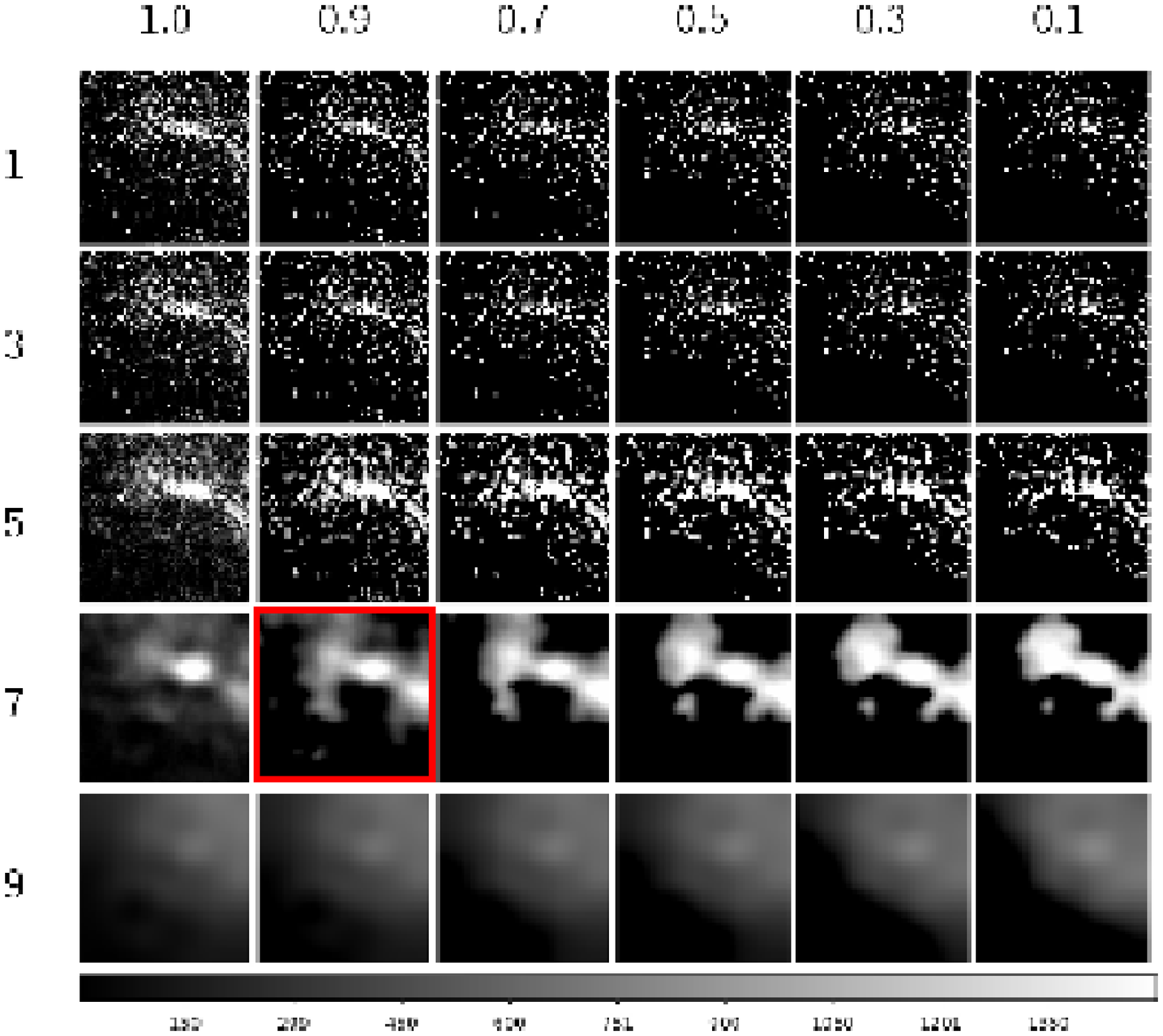}
 \end{center}
\caption{The similar figure with Figure \ref{fig:double_star_mu_beta} for Cas A.
}\label{fig:casa_mu_beta}
\end{figure}

We test the performance of our proposed reconstruction method by using simulated observed images.
To make the simulated images and perform the image reconstruction,
we use the response of the detector $t(v, u)$.
It is made by a ray-tracing code ``xrtraytrace'' \citep{Yaqoob+2018}.
The optics parameter of an {\it Hitomi} SXT-like light-weight telescope \citep{Okajma+2016}
is adopted in the ray-tracing.
The angular resolution we set for this ray-tracing is about 1 arcmin.
The off-focus images through the double masks are calculated
by offsets with a 2 arcsec pitch. See \citet{Maeda+2018} for details.

We simulate images of double point sources with a close separation of 4 arcsec
and diffuse sources such as Cas A and Crab Nebula.
The input images are shown in Figure \ref{fig:reconst} (a, e, i) and
the simulated observed images on the off-focus imager are the panels (b, f, j), respectively.
Both image sizes of the off-focus imager and the reconstructed sky image are $60 \times 60$ pixels.
We implemented the algorithm with C++, utilizing BLAS library.

Figures \ref{fig:double_star_mu_beta} and \ref{fig:casa_mu_beta} show
the reconstructed images of the double point sources and Cas A
along with the hyper parameters varied.
The best reconstructed image with the best hyper parameters
is determined by the ten-fold cross-validation and shown as the marked panel
in these figures.
Here, we search the parameter among 121 pairs of ($\beta$, $\mu$):
$\beta$ is varied from $0.1$ to $1.1$ by a step of $0.1$,
and $\log_{10} \mu$ is varied from $0$ to $10$ by a step of $1$.
The right two columns of Figure \ref{fig:reconst} presents the reconstructed images
for the double stars, Crab Nebula and Cas A obtained by the Richardson-Lucy method and our proposed method.

To estimate the angular resolution of our method,
we evaluate the successful separation rate for two close point sources with
an angular distance of 4 arcsec as shown in Figure \ref{fig:reconst} (a) for 100 simulations.
In the cases of total photons with $10^4$, $10^3$ and $10^2$,
the success rates are $100\%$ (100 out of 100 trials), 
$99\%$ (99/100) and $12\%$ (12/100), respectively.

The processing speed is measured by using a budget computer equipped with 
Intel Core-i7 with 3.20 GHz.
The image with $60 \times 60$ pixels are reconstructed by 70 sec on average for each hyper-parameter.

\section{Discussion and conclusion}

By the reconstruction of double stars with close separation (Figure \ref{fig:reconst}, d),
we demonstrated that the booster system with our image reconstruction method
can achieve superb resolution of 4 arcsec,
if photon counts from a point source is more than about $5 \times 10^2$.

Although the reconstructed images vary with the parameters ($\beta$, $\mu$) 
as shown in Figures \ref{fig:double_star_mu_beta} and \ref{fig:casa_mu_beta},
the best parameters can be determined by the cross-validation, and
the best image looks plausible.
For point sources, small smoothness parameter is selected
by the cross-validation (Figure \ref{fig:double_star_mu_beta}), 
while for diffuse objects, large smoothness parameters are selected.
As shown in Figure \ref{fig:reconst}, the images obtained by the Richardson-Lucy method are noisy,
while the images obtained by our method are less noisy and look similar to the input images.

As shown in the second paragraph of Section 2, the total intensity of the X-ray image and
the pattern of image are estimated separately. Then, the total intensity is simply estimated
by $S^*$ and one sigma error is by the square root of $S^*$.
The reconstructed pattern of images depend
on the balance between photon statistics and complexity of input images.
Roughly speaking, complex images require larger photon statistics than
images containing sparse point sources do.
Quantitative estimation of necessary number of photons
for the reconstruction will be a future work.
In this paper, the energy dependency of an X-ray image is not taken into consideration.
The reconstruction of image and energy spectrum at once would be the next challenge,
e.g. by introducing a regularization of smoothness for energy channel.

In this paper, we have successfully demonstrated
that 1-2 order of magnitude improvement of angular resolution
can be realized by using angular resolution booster combined with the image reconstruction.
This system needs only a simple thus possibly low-cost modification
for the current well-established telescopes.
Thus, our method will progress X-ray astronomy in near future.

%



\begin{ack}
This research is supported by CREST, the Japan Science and Technology Agency (JST), JPMJCR1414,
and in part by JSPS Grants-in-Aid for Scientific Research (KAKENHI) Grant No. 17K05395.
YM gratefully acknowledges funding from the Tanaka Kikinzoku Memorial Foundation.
\end{ack}

\appendix 

\section*{Solution of the sub-problem}

An iteration step of optimization in the proximal gradient method (Equation \ref{eq: proxmap})
is solved by minimizing the following Lagrangian with a Lagrange multiplier $\tau$.
\begin{eqnarray}
  l(\rho, \tau) & = & \sum_u \left[ - [m_u^{(r)} - (1 - \beta)] \log \rho(u)
    + \frac{L_k}{2} \left(\rho_u - \sigma_u^{(k)} \right)^2 \right]
  + \tau \left( \sum_u \rho_u- 1 \right).
\end{eqnarray}
By fixing $\tau$, it is solved in a closed form:
\begin{equation}
  \rho_u(\tau) = \left\{ \begin{array}{ll}
    \frac{1}{2} \left[ b_u^{(k)}(\tau) + \sqrt{ D_u^{(k)}(\tau) } \right] &
    (m_u^{(r)} - (1 - \beta) > 0) \\
    b_u^{(k)}(\tau) & (m_u^{(r)} - (1 - \beta) = 0) \\
    0 & (m_u^{(r)} - (1 - \beta) < 0)
  \end{array} \right. ,
\end{equation}
where
\begin{eqnarray}
  D_u^{(k)}(\tau) & = & {b_u^{(k)}(\tau)}^2 + 4 \frac{m_u^{(r)} - (1 - \beta)}{L_k}, \\
  b_u^{(k)}(\tau) & = & \sigma_u^{(k)} - \frac{\tau}{L_k}.
\end{eqnarray}

Here, the $\tau$ must satisfy the condition $s(\tau) = \sum_u \rho_u(\tau) - 1 = 0$.
Since the function $s(\tau)$ is convex, the solution can be obtained by the Newton-Raphson method.
The updating rule of the $i$-th Newton-Raphson step is
\begin{equation}
  \tau^{(i + 1)} = \tau^{(i)} - \frac{s(\tau^{(i)})}{s^\prime(\tau^{(i)})},
\end{equation}
where 
\begin{eqnarray}
  s^\prime(\tau) & = & \sum_u \rho_u^\prime(\tau), \\
  \rho_u^\prime(\tau) & = & \left\{ \begin{array}{ll}
    - \frac{1}{2 L_k} \frac{ b_u^{(k)}(\tau) + \sqrt{ D_u^{(k)}(\tau) } }{\sqrt{ D_u^{(k)}(\tau) }}
    & (m_u^{(r)} - (1 - \beta) > 0) \\
    -\frac{1}{L_k} & (m_u^{(r)} - (1 - \beta) = 0) \\
    0 & (m_u^{(r)} - (1 - \beta) < 0)
  \end{array} \right. .
\end{eqnarray}


\end{document}